\documentclass[a4paper,
               keeplastbox,   
               ]{jacow}
\usepackage{pdfpages,multirow,ragged2e} %
\makeatletter%
    \ifboolexpr{bool{xetex}}
     {\renewcommand{\Gin@extensions}{.pdf,%
                        .png,.jpg,.bmp,.pict,.tif,.psd,.mac,.sga,.tga,.gif,%
                        .eps,.ps,%
                        }}{}
\makeatother

\ifboolexpr{bool{xetex} or bool{luatex}} 
 {}                                      
 {\usepackage[utf8]{inputenc}}           

\usepackage[USenglish]{babel}

\ifboolexpr{bool{jacowbiblatex}}%
 {%
  \addbibresource{jacow-test.bib}
  \addbibresource{biblatex-examples.bib}
 }{}
\begin{document}

\title{Linac\_Gen: integrating machine learning and particle-in-cell methods for enhanced beam dynamics at Fermilab}

\author{Abhishek Pathak\thanks{abhishek@fnal.gov\\This manuscript has been authored by Fermi Research Alliance, LLC under Contract No. DE-AC02-07CH11359 with the US Department of Energy, Office of Science, Office of High Energy Physics.}, Fermi  National Accelerator Laboratory, Batavia, USA\\ 
		}
	
\maketitle

\begin{abstract}
Here, we introduce \texttt{Linac\_Gen}, a tool developed at Fermilab, which combines machine learning algorithms with Particle-in-Cell methods to advance beam dynamics in linacs. \texttt{Linac\_Gen} employs techniques such as Random Forest, Genetic Algorithms, Support Vector Machines, and Neural Networks, achieving a tenfold increase in speed for phase-space matching in linacs over traditional methods through the use of genetic algorithms. Crucially, \texttt{Linac\_Gen}'s adept handling of 3D field maps elevates the precision and realism in simulating beam instabilities and resonances, marking a key advancement in the field. Benchmarked against established codes, \texttt{Linac\_Gen} demonstrates not only improved efficiency and precision in beam dynamics studies but also in the design and optimization of linac systems, as evidenced in its application to Fermilab's PIP-II linac project. This work represents a notable advancement in accelerator physics, marrying ML with PIC methods to set new standards for efficiency and accuracy in accelerator design and research. \texttt{Linac\_Gen} exemplifies a novel approach in accelerator technology, offering substantial improvements in both theoretical and practical aspects of beam dynamics.
\end{abstract}

\section{INTRODUCTION}

Modeling of an accelerator plays a pivotal role in the successful realization of the system, encompassing a multifaceted scope from the design of individual components, such as cavities and magnets, to the validation of these designs within an accelerating lattice, and the study of beam dynamics within this lattice.

The indispensable nature of these simulation tools has driven substantial global efforts, resulting in the development of numerous simulation tools tailored to meet diverse needs. These needs range from designing various structures to studying various physical phenomena occurring within an accelerator, such as acceleration, focusing, resonances, beam instabilities, and beam losses through different mechanisms.

With the forthcoming PIP-II\cite{pip2-optimization,cdr} upgrade at Fermilab, which operates at a beam current of 2 mA within the space-charge dominated regime, a comprehensive simulation tool was required. This tool needed to optimize longitudinal dynamics by considering synchronous particle dynamics, enhance transverse focusing properties of the lattice to avoid parametric resonances, perform beam matching along the linac, and execute fully three-dimensional, relativistic particle tracking simulations with space charge to produce results as close to real-world scenarios as possible. Additionally, the tool needed to offer flexibility for code modification, allowing for the integration of new ideas and adjustments to existing schemes. Lastly, the tool needed to be efficient enough to minimize the time required for numerous runs during linac and beam tuning procedures.

Given the lack of an existing tool that met all these requirements, we developed a new simulation tool named \texttt{Linac\_Gen}. This code is organized into five major modules. The first module designs the longitudinal lattice and performs longitudinal optimization based on the required energy gain and the types of cavities used. The second module utilizes the optimized longitudinal lattice to perform transverse lattice optimization, focusing on the point of operation in the Hoffmann stability chart with space charge. The third module employs machine learning techniques, such as genetic algorithms, to achieve inter-cryomodule matching throughout the lattice. The fourth module conducts fully three-dimensional Particle-in-Cell simulations to track the beam within the lattice. Finally, the fifth module applies reinforcement learning to further optimize the lattice, aiming to enhance beam quality along the lattice using the machine learning module.

\section{Longitudinal Optimization}

Once the types of cavities to be used, along with their respective $\beta_g$ for different sections of the accelerator, have been decided, the first step is to perform single-particle dynamics. This step involves determining the accelerating gradient, synchronous phase, and consequently the longitudinal phase advance while considering the limitations of the cavities under consideration. The following constraints were imposed while optimizing the longitudinal lattice:

\begin{itemize}
    \item \textbf{Adiabatic Control of Phase Advance:} Maintain adiabatic variation of phase advance per unit length ($k\zeta_0$) to avoid large amplitude oscillations and beam size growth.
    \item \textbf{Structure Phase Advance Variation:} Keep zero current phase advance per period ($\sigma\zeta_0$) below 90° to prevent envelope instabilities.
    \item \textbf{Synchronous Phase:} Ensure the synchronous phase remains within limits to maintain stable longitudinal motion.
    \item \textbf{Energy Gain per Cavity:} Optimize the energy gain per cavity to match the desired energy profile along the accelerator.
    \item \textbf{Cavity Limitations:} Consider the operational limits of the cavities, such as maximum achievable gradient and power constraints.
\end{itemize}

By adhering to these constraints, we aim to achieve an optimized longitudinal lattice that ensures stable beam dynamics and meets the design requirements of the accelerator.
\section{Transverse Lattice Optimization}

The optimization of the transverse lattice involves several critical steps to ensure stable and efficient beam dynamics. The following steps are undertaken:

\begin{itemize}
    \item \textbf{Transverse Focusing Dynamics:} Account for the contributions of RF cavity electromagnetic fields to $k^2\zeta_0(s, \gamma)$, which affect betatron oscillations. Prevent parametric resonance\cite{sc} by ensuring $\sigma t_0$ does not equal an integer multiple of $\sigma z_0/2$.
    \item \textbf{Mitigating Space Charge Instabilities:} Position lattice footprints in resonance-free zones on the Hofmann diagram to prevent emittance exchange. Ensure that the tune depression is maintained above 0.5 to maintain minimum energy exchange rate.
\end{itemize}

By implementing these steps, we aim to achieve an optimized transverse lattice that ensures stable transverse dynamics, prevents resonances, and mitigates space charge instabilities.
\section{Beam Matching}

In an accelerator, six-dimensional phase space matching is essential throughout the accelerator to avoid beam envelope instability and Landau damping. During this process, the beam Twiss parameters at the exit of a lattice period should match those at the entrance of the period. This matching is ensured by varying the transverse focusing fields, accelerating gradient, and phase.

To achieve this, we adopted a genetic algorithm for optimization. The genetic algorithm operates as follows:

\begin{enumerate}
    \item \textbf{Initialization:} Generate an initial population of candidate solutions, where each candidate is a set of parameters defining the transverse focusing fields, accelerating gradient, and phase.
    \item \textbf{Selection:} Evaluate the fitness of each candidate solution based on a predefined objective function. Select a subset of the best-performing candidates for reproduction.
    \item \textbf{Crossover:} Combine pairs of selected candidates to produce offspring by exchanging portions of their parameter sets.
    \item \textbf{Mutation:} Introduce small random changes to the offspring parameter sets to maintain genetic diversity.
    \item \textbf{Iteration:} Repeat the selection, crossover, and mutation steps for several generations until convergence criteria are met.
\end{enumerate}

The objective function \( f(x) \) used in our genetic algorithm is designed to minimize the difference between the Twiss parameters at the exit and entrance of a lattice period, and is given by:
\[
f(x) = \sum_{i=1}^{N} \left( \alpha_{i,\text{exit}} - \alpha_{i,\text{entrance}} \right)^2 + \left( \beta_{i,\text{exit}} - \beta_{i,\text{entrance}} \right)^2 + 
\]
\[
\left( \gamma_{i,\text{exit}} - \gamma_{i,\text{entrance}} \right)^2
\]
where \( \alpha_i \), \( \beta_i \), and \( \gamma_i \) are the Twiss parameters for the \( i \)-th dimension of the phase space, and \( N \) is the total number of dimensions.

By implementing this genetic algorithm, we ensure that the beam parameters remain consistent throughout the accelerator, leading to a stable and optimized beam matching process.
\section{Implementation of the Particle-In-Cell Algorithm for Particle Tracking with Space Charge}

The implementation of the Particle-In-Cell (PIC) algorithm for particle tracking with space charge involves several standard steps to accurately simulate beam dynamics:

\begin{itemize}
    \item \textbf{Covariant Beam Matrix and Random Number Generator:} Used to generate the initial particle distribution, ensuring it accurately reflects the desired beam properties.
    \item \textbf{Cloud-In-Cell (CIC) for Charge Deposition:} Assigns charges to the nearest grid points, effectively distributing particle charge over the grid.
    \item \textbf{Poisson Solver with Successive Over Relaxation (SOR):} Solves the Poisson equation to obtain the electrostatic potential efficiently and quickly.
    \item \textbf{Boris Pusher for Particle Pushing:} Updates positions and velocities of particles accurately and stably in the presence of electromagnetic fields.
    \item \textbf{Linear 3D Interpolator for Intermediate Interpolations:} Ensures smooth and accurate interpolation of field values at particle positions.
\end{itemize}

By integrating these steps, the PIC algorithm provides a robust and accurate framework for simulating particle tracking with space charge effects in accelerators.
\section{Validation Study}

\begin{figure}[htbp]
    \centering
    \includegraphics[width=0.8\linewidth]{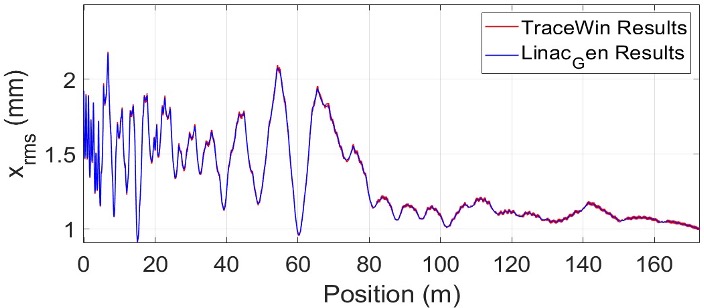}
    \caption{RMS envelope comparison for TraceWin and \texttt{Linac\_Gen} simulations.}
    \label{fig:rms_envelope}
\end{figure}

To validate the performance of our code \texttt{Linac\_Gen}, we conducted a comparative study with the commercial simulation tool TraceWin. Figure~\ref{fig:rms_envelope} shows the RMS envelope comparison for the simulations performed with TraceWin and \texttt{Linac\_Gen}. The results show agreement within three percent, indicating the accuracy of our code.

In addition to employing genetic algorithms for beam matching, we have incorporated several other machine learning algorithms to further enhance the performance and accuracy of our simulations. These include:

\begin{itemize}
    \item \textbf{Convolutional Neural Networks (CNNs):} Used for analyzing phase space images to optimize beam profiles, leveraging their high accuracy in spatial pattern recognition, which is essential for beam shaping.
    \item \textbf{Long Short-Term Memory Networks (LSTMs):} Employed to model time-dependent beam dynamics and predict future beam states. LSTMs are particularly useful for capturing long-term dependencies in temporal data, crucial for dynamic stability analysis.
    \item \textbf{Random Forests:} Utilized to predict linac operational parameters from historical data. This algorithm handles high-dimensional spaces effectively, making it important for parameter tuning.
    \item \textbf{K-means Clustering:} Applied for the unsupervised categorization of beam states to detect anomalies. This method efficiently identifies distinct operational regimes, which is beneficial for diagnostics.
\end{itemize}

\begin{figure}[htbp]
    \centering
    \includegraphics[width=0.8\linewidth]{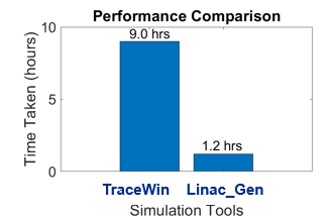}
    \caption{Performance comparison between TraceWin and \texttt{Linac\_Gen}.}
    \label{fig:performance_comparison}
\end{figure}

By utilizing these advanced machine learning techniques, the time required for the simulation has been reduced significantly. Figure~\ref{fig:performance_comparison} demonstrates the performance comparison between TraceWin and \texttt{Linac\_Gen}, highlighting a nine-fold decrease in simulation time.

The integration of genetic algorithms, reinforcement learning, and other machine learning techniques in \texttt{Linac\_Gen} has not only improved the simulation accuracy but also significantly reduced the computational time, making it a highly efficient tool for accelerator modeling and optimization.

\section{Conclusion}

In this paper, we introduced \texttt{Linac\_Gen}, a novel simulation tool developed at Fermilab, integrating machine learning algorithms with Particle-in-Cell (PIC) methods to advance beam dynamics in linear accelerators. The tool leverages genetic algorithms, convolutional neural networks, long short-term memory networks, random forests, and K-means clustering to optimize various aspects of linac design and operation.

The longitudinal and transverse lattice optimization processes were detailed, highlighting the control of phase advance, mitigation of space charge instabilities, and the use of genetic algorithms for six-dimensional phase space matching. Furthermore, the implementation of the PIC algorithm was described, incorporating steps such as charge deposition using the Cloud-In-Cell method, solving the Poisson equation with successive over relaxation, and particle pushing using the Boris algorithm.

Validation studies demonstrated that \texttt{Linac\_Gen} achieves high accuracy in beam dynamics simulations, with results agreeing within three percent of those obtained using TraceWin. Additionally, the use of advanced machine learning techniques significantly reduced simulation times, achieving a nine-fold decrease compared to traditional methods.

The integration of these sophisticated algorithms not only enhances the precision and efficiency of simulations but also facilitates the design and optimization of linac systems. \texttt{Linac\_Gen} sets new standards for accelerator modeling, offering substantial improvements in both theoretical and practical aspects of beam dynamics, and exemplifying a novel approach in lattice design and simulation technology.

%
%
\ifboolexpr{bool{jacowbiblatex}}%
    {\printbibliography}%
    {%
    

} 
\end{document}